\documentclass[prx,twocolumn,superscriptaddress]{revtex4-2}

\usepackage[english]{babel}
\usepackage[utf8]{inputenc}
\usepackage[colorinlistoftodos, color=green!40, prependcaption]{todonotes}
\usepackage[pdftex, pdftitle={Article}, pdfauthor={Author}]{hyperref} 
\usepackage{amssymb}
\usepackage{braket}
\usepackage{comment}
\usepackage{ragged2e}
\usepackage{subcaption}
\usepackage{tikz}
\usetikzlibrary{arrows.meta, positioning}
\usepackage{graphicx}
\newcommand{\e}{\text{e}}
\newcommand{\dx}{\text{d}}

\usepackage{amsthm}
\usepackage{mathtools}
\usepackage{physics}
\usepackage{caption}
\usepackage{xcolor}
\usepackage{graphicx}
\usepackage[left=23mm,right=13mm,top=35mm,columnsep=15pt]{geometry} 
\usepackage{adjustbox}
\usepackage{placeins}
\usepackage[T1]{fontenc}
\usepackage{lipsum}
\usepackage{csquotes}

\usepackage{graphicx}
\usepackage{dcolumn}
\usepackage{bm}

\usepackage[utf8]{inputenc}
\usepackage[T1]{fontenc}
\usepackage{etoolbox}

\makeatletter
\def\@email#1#2{%
 \endgroup
 \patchcmd{\titleblock@produce}
  {\frontmatter@RRAPformat}
  {\frontmatter@RRAPformat{\produce@RRAP{*#1\href{mailto:#2}{#2}}}\frontmatter@RRAPformat}
  {}{}
}%
\makeatother
\begin{document}

\preprint{AIP/123-QED}

\title[The Effects of Permanent Dipoles on Dark States in Molecular Dimers]{The Effect of Permanent Dipoles on Dark States in Molecular Dimers}
\author{Matthew Freed}
	\email[Correspondance email: ]{m.freed@surrey.ac.uk}
	\affiliation{School of Mathematics and Physics
 ,  University of Surrey, Guildford, GU2 7XH, United Kingdom}

\author{ Dominic M. Rouse}
	\affiliation{School of Physics and Astronomy, University of Glasgow, Glasgow G12 8QQ, Scotland, United Kingdom}
	\affiliation{Department of Physics and Astronomy, University of Manchester, Oxford Road, Manchester M13 9PL, United Kingdom}

    \author{Andrea Rocco}
	\affiliation{School of Biosciences,  University of Surrey, Guildford, GU2 7XH, United Kingdom}
	\affiliation{School of Mathematics and Physics
 ,  University of Surrey, Guildford, GU2 7XH, United Kingdom}
\author{Jim Al-Khalili}
	
	\affiliation{School of Mathematics and Physics
 ,  University of Surrey, Guildford, GU2 7XH, United Kingdom}
	
\author{Marian Florescu}
	\affiliation{Optoelectronics Research Centre, University of Southampton, Southampton SO17 1BJ, United Kingdom}
    \author{Adam Burgess}%
\affiliation{SUPA, Institute of Photonics and Quantum Sciences, Heriot-Watt University, Edinburgh, EH14 4AS, UK}

\date{\today} 

\begin{abstract}
Many organic molecules possess large permanent dipole moments that differ depending on the electronic state. These permanent dipoles influence both intermolecular coupling and interactions with the optical fields, yet they are often neglected in typical theoretical quantum optics treatments. Here, we investigate the optical properties and their effect on dark states of dimers possessing such permanent dipoles. We show that when monomers have excitation-dependent permanent dipoles, optical transitions between the bright and dark states of the dimer are enabled. We investigate how permanent dipoles allow for the existence of static driving terms between the ground and excited states of each monomer. In turn, these can cause the excited states of the monomers to couple indirectly to the zero excitation state of the dimer. This leads to interference between permanent and transition dipoles and can result in the formation of dark states that are entirely localised. Furthermore, dark states formed through indirect coupling exhibit enhanced robustness against energy level fluctuations, which may improve the efficiency of the design of photovoltaic devices.
\end{abstract}

\keywords{permanent dipole, dark state, dimer}

\maketitle

\section{Introduction}
The study of light-matter interactions in molecular systems has attracted considerable attention for their promising role in energy capture and transfer, particularly in the context of light-harvesting systems~\cite{rouse_optimal_2019,riede_efficient_2011, ameri_highly_2013, mancal_exciton_2010, chan_single-photon_2018}. These systems operate by capturing photon energy through the interaction between the optical field and the transition dipoles of a molecular system, forming an exciton. When multiple monomers - individual molecular subunits - are considered, the excitons may become delocalised, their eigenstates distributed across multiple spatial sites, and they can act as mobile energy carriers, mediating transfer over nanometer scales on ultrafast timescales. The properties of the excitons, such as their recombination rate, are strongly determined by the dipoles of the monomers and their interactions. To optimise the efficiency of such systems, it is important to ensure that the exciton energy is not lost either via vibrational relaxation or through the emission of a photon\cite{ghosh_decoupling_2024,merrifield_theory_1964}. 

Optical losses, on which we focus in this paper, can be greatly reduced by the formation of dark states, coherent superpositions of excited monomer states that decoupled from the optical field~\cite{zhang_dark_2016, fruchtman_photocell_2016, zhang_delocalized_2015, mondaini_dark_2018, hu_double-excitation_2018}. Dark states are formed when the transition dipoles of the participating monomers destructively interfere. Although these states are unable to absorb light directly, they can significantly enhance energy transfer by acting as intermediate states that suppress radiative losses, thereby improving overall efficiency~\cite{davidson_eliminating_2022,davidson_dark_2020}. As the total dipole strength across the monomers is conserved, it is typical that whenever dark states form, so too do bright states. These are states where the transition dipoles constructively interfere, leading to absorption and emission rates that can exceed the sum of the composite parts of the combined system. This can improve the rate of energy capture~\cite{creatore_efficient_2013}, but can also contribute to a lower transport efficiency of the system.

Although there have been extensive studies of excitonic coupling and dark states, the role of permanent dipoles has often been neglected~\cite{garziano_one_2016, young_mixed_2020, hestand_expanded_2018,stokes_master_2018, rouse_avoiding_2021, breuer_theory_2007}, despite many organic molecules exhibiting significant permanent dipole moments, (often several Debye in magnitude)~\cite{drobizhev_resonance_2007, alaraby_salem_two-photon_2015, deiglmayr_permanent_2010, rogers_investigating_2006, guerout_ground_2010, jagatap_contributions_2002}. In monomeric systems, these dipoles have been shown to induce multi-photon processes and even suppress optical activity under certain driving conditions~\cite{burgess_strong_2023, mirzac_microwave_2021, mandal_polarized_2020, shim_permanent_1999, chestnov_ensemble_2016, anton_radiation_2016, chestnov_terahertz_2017, warburton_giant_2002}. In a dimer, permanent dipoles can interact with each other and with transition dipoles via static Coulomb forces, shifting the eigenstates of the monomers and modifying the interference conditions defining bright and dark state formation. Additionally, the permanent dipoles of the monomers can interfere with the transition dipoles. These previously unaccounted for effects could be harnessed to tune the radiative lifetimes of both dark and bright states, change the spatial localisation of excitons, and optimise quantum energy transfer pathways. Here, we aim to highlight the importance of permanent dipoles in the optical properties of organic molecules and the potential advantages they can bring to exciton transport. We focus on dimeric systems in order to keep the analysis tractable and act as a building block for more complex systems.

In this paper, we investigate the effects that the permanent dipoles of monomers have on the transition rates between the eigenstates of the dimer. We focus our efforts on structures where the eigenbasis can be expressed analytically in terms of the properties of the monomer. These are characterised by whether the Coulomb interaction causes the excited states of the monomers to be directly coupled to each other or whether they couple through an intermediate state. We find that when the monomers' excited states couple through their shared ground state, the permanent dipoles can interfere with the transition dipoles, allowing dark states to form that are more robust to fluctuations on the system Hamiltonian.

The remainder of the paper is structured as follows. In Sec.\ref{sec:Model}, we present the Hamiltonian for a molecular dimer in which each monomer possesses both transition dipoles and state-dependent permanent dipoles. The transition rates for the dimer are calculated through the secularised Redfield master equation, and the conditions for the formation of a dark state are identified. In Sec.\ref{Sec: Rates}, we investigate the dimer systems where the monomers' excited states are coupled directly, through the ground state and a combination of both, deriving the transition rates and conditions for dark states for each case. Sec.\ref{Sec: Robust} investigates how the robustness of the dark states is affected by the type of coupling being used. We conclude in Sec.~\ref{Sec: Conc} with a summary and outlook.

\section{The Model} \label{sec:Model}

We consider a molecular dimer composed of two monomeric sub-units, which we restrict to their ground- and first-excited electronic states. To explicitly capture the contribution of each monomer to the system dynamics, we construct the Hamiltonian from the isolated properties of the monomers and their relative position and orientation. These include the bare excitation energies and electric dipole moments of the monomers. Unlike many prior models, we explicitly include the permanent dipole moments in both the ground and excited states that arise from asymmetries in the monomer. Throughout this work, all spatial vectors are written in bold, while system operators are represented using either hats or bra-ket notation, depending on the context. We also use natural units where \( \hbar = c = \epsilon^{-1} = 1 \) throughout unless otherwise specified.

\subsection{Hamiltonian}

\begin{figure}[ht]
    \centering
    \begin{minipage}{0.45\textwidth}
        \centering
        \begin{tikzpicture}[>=Stealth, thick, scale=1]

\def\Ezero{0}
\def\Eone{1.9}
\def\Etwo{2}
\def\Ethree{\Eone+\Etwo} 

\def\xleft{0}
\def\xmid{1.2}
\def\xright{2.4}

\draw[thick] (\xmid-0.5,\Ezero) -- (\xmid+0.5,\Ezero) node[right] {$E_0$};

\draw[thick] (\xleft-0.5,\Eone) -- (\xleft+0.5,\Eone) node[right] {$E_1$};

\draw[thick] (\xright-0.5,\Etwo) -- (\xright+0.5,\Etwo) node[right] {$E_2$};

\draw[thick] (\xmid-0.5,\Ethree) -- (\xmid+0.5,\Ethree) node[right] {$E_3$};

\node[left] at (\xmid-0.5,\Ezero) {$\ket{0}$};
\node[left] at (\xleft-0.5,\Eone) {$\ket{1}$};
\node[left] at (\xright-0.5,\Etwo) {$\ket{2}$};
\node[left] at (\xmid-0.5,\Ethree) {$\ket{3}$};

\draw[<->, thick, black!50!white] (\xmid-0.1,\Ezero) -- (\xleft,\Eone);
\draw[<->, thick, white!20!black] (\xmid+0.1,\Ezero) -- (\xright,\Etwo);
\draw[<->, thick, white!20!black] (\xleft,\Eone) -- (\xmid-0.1,\Ethree);
\draw[<->, thick, black!50!white] (\xright,\Etwo) -- (\xmid+0.1,\Ethree);
\node [label={[label distance=3.5cm]100:(a)}] {};

\end{tikzpicture}
        
    \end{minipage}%
    \hfill

    \begin{minipage}{0.45\textwidth}
        \centering
        
\begin{tikzpicture}[>=Stealth, thick, scale=1]

\def\sep{4}

\draw[fill=black!10] (0,0) circle (0.5);
\node at (0,-0.9) {Monomer 1};

\draw[->, black!50!white, thick] (0,0) -- (1.2,0.7) node[above right] {\pmb{$G_1$}};
\draw[->, black!50!white, thick] (0,0) -- (1.2,-0.4) node[below right] {\pmb{$E_1$}};
\draw[->, black!50!white, thick] (0,0) -- (-0.2,1.4) node[above] {\pmb{$\mu_1$}};

\draw[fill=black!20] (\sep,0) circle (0.5);
\node at (\sep,-0.9) {Monomer 2};

\draw[->, white!20!black, thick] (\sep,0) -- ({\sep+1.1},0.3) node[above right] {\pmb{$G_2$}};
\draw[->, white!20!black, thick] (\sep,0) -- ({\sep+1.1},-0.8) node[below right] {\pmb{$E_2$}};
\draw[->, white!20!black, thick] (\sep,0) -- (\sep+0.2,1.4) node[above] {\pmb{$\mu_2$}};

\draw[->, thick, black, dashed] (0.5,0) -- (\sep-0.5,0) node[midway, above] {\(\mathbf{R}_2 - \mathbf{R}_1\)};
\node [label={[label distance=1.4cm]120:(b)}] {};
\end{tikzpicture}
        
    \end{minipage}%
\caption{\justifying (a) Energy level structure in the site basis without any electostatic interactions. Here, the lighter and darker arrows represent transitions that occur through \(\pmb{\mu}_1\) and \(\pmb{\mu_2}\) respectively. (b) Schematic of dipole orientations of each monomer and there relative positions.}\label{fig1}
\end{figure}
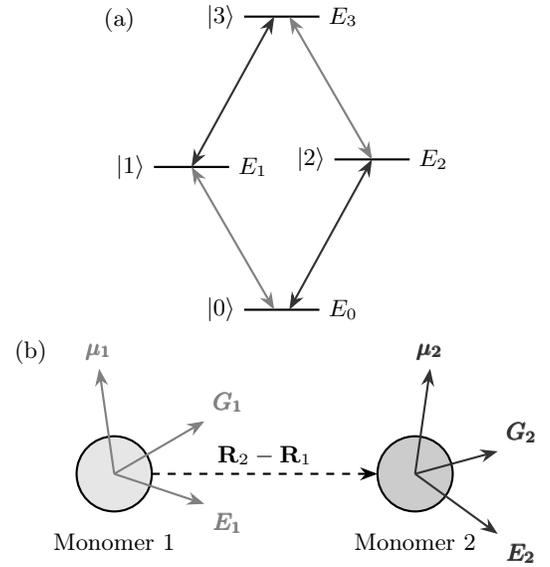

We define the total Hamiltonian in the site basis to describe the electronic configuration of two monomers labelled 1 and 2. In this basis, the state \( \ket{0} \) denotes the configuration in which both monomers are in their ground states. A single excitation in monomer \(1\) or \(2\) is represented by \(\ket{1}\) or \(\ket{2}\) respectively. To fully define the Hamiltonian, we also include the doubly excited state \( \ket{3} \), where both monomers are excited; however, most of our analysis will focus on the zero and single-excitation manifolds. Throughout the paper, the indices \(i,j \in \{1,2\}\) refer to the monomers, while \(a,b \in \{0,1,2,3\}\) are used for the states of the system. The Hamiltonian of the dimer system in the site basis is
\begin{align} \label{Eq: H0 lab} 
    \hat{H}_S = \sum_{a=1}^3 E_a \ket{a}\bra{a} + \hat{V}_0, 
\end{align}
where \( E_a \) is the excitation energy of the respective monomers when isolated for \( a = 1,2 \), and \( E_3 = E_1 + E_2 \). A schematic of this is shown in Fig(\ref{fig1}a). Each term accounts for the kinetic and potential energies of each monomer's electronic Hamiltonian when they are isolated. The electrostatic interaction operator, \( \hat{V}_0 \), then accounts for the electrostatic interactions between each monomer due to their charge distribution. This can be expressed in the site basis by defining \(\hat{V}_{0} = \sum_{a} \sum_b Q_{ab}\ket{a}\bra{b}\), where \(Q_{ab}\) is the energy of the electrostatic interactions between each monomer in their respective states. In general, this energy is defined as
\begin{align}\label{Eq: Coulomb Gen}
    Q_{ab} = \bra{a}\sum_{\alpha} \sum_\beta \frac{q_{\alpha_1} q_{\beta_2}}{4 \pi  |\hat{\mathbf{r}}_{\alpha_1}-\hat{\mathbf{r}}_{\beta_2}|}\ket{b},
\end{align}
where \( q_{\alpha_j} \) and \( \hat{\mathbf{r}}_{\alpha_j} \) are the charge and position operators of the constituent particle \( \alpha \) in monomer \( j \), respectively. It is possible to model the electrostatic coupling directly in terms of the dipoles of each monomer using the point-dipole approximation. However, this can become inaccurate when the monomers are in close proximity to each other \cite{tretiak_exciton_2000}. When using the point-dipole approximation as a rough guideline, it becomes apparent that the presence of permanent dipoles allows any coupling term in \(\hat{V}_0\) to be significant, where otherwise the inter excited-state coupling terms, \(Q_{12}\) and \(Q_{21}\), dominate. Since the true values of \(Q_{ab}\) are difficult to solve in general and depend on many features of the molecular wavefunctions \cite{fujimoto_electronic_2021}, we treat the values of \(Q_{ab}\) as independent parameters, while observing that they are often correlated with the dipoles of the monomer.  

Both monomers also couple to the optical field, which we decompose into its modes labelled \( k \), each with its respective energy \( \nu_k \), momentum \(\textbf{k}\),  and perpendicular unit polarization vector \( \mathbf{e}_k \). The free Hamiltonian of this optical field is the standard \(\hat{H}_O = \sum_k \nu_k \hat{a}_k^\dagger \hat{a}_k,\)
where \( \hat{a}_k \) and \( \hat{a}_k^\dagger \) are the annihilation and creation operators of the \(k^\text{th}\) mode. As the wavelengths of the relevant photon modes are sufficiently larger than the size of each monomer, corresponding to approximately 1eV$\sim $700nm, the electric dipole approximation is made when coupling each monomer to the optical field. 

In the multipolar gauge, the interaction between each monomer and the optical field is described in terms of its dipole,
where \(\hat{\mathbf{d}}_j\) is the dipole operator for the \( j^\text{th} \) monomer defined as \(
\hat{\mathbf{d}}_j = \sum_{\alpha} q_{\alpha_j} \hat{\mathbf{r}}_{\alpha_j}\). When projected onto the site basis, this operator can be expressed in terms of spatial vectors and dimer state projectors \( \ket{i}\bra{j} \). For \( i \neq j \), the dipole operator of monomer \(j\) is
\begin{multline}
    \hat{\mathbf{d}}_j = (\ket{0}\bra{0} + \ket{i}\bra{i}) \pmb{G}_j + (\ket{j}\bra{j} + \ket{3}\bra{3}) \pmb{E}_j\\
    +\left((\ket{0}\bra{j} + \ket{i}\bra{3})+h.c.\right)\pmb{\mu}_j,
\end{multline}
where \( \pmb{G}_j \) and \( \pmb{E}_j \) are the permanent dipole moments of the ground and excited states of monomer \( j \), respectively. A schematic of this is shown in Fig.(\ref{fig1}b). The transition dipole moment \( \pmb{\mu}_j \) is generally complex, but we define the site basis such that it is always real. For convenience, we also define the change in the permanent dipole as \(\pmb{\Delta}_j = \pmb{E}_j - \pmb{G}_j \) as this is the main form through which the permanent dipoles interact with the optical field.\\

The interaction Hamiltonian couples these dipoles to the transverse electric field. This Hamiltonian can be written as
\begin{align}
    \hat{H}_I = \sum_k \sum_{j=1}^2 -i (\hat{\mathbf{d}}_j\cdot \mathbf{e}_k) f_k \left(  \hat{a}^\dagger_k\text{e}^{i \textbf{k}\cdot \textbf{X}_j } -h.c.\right),
\end{align}
where \(f_k\) is the real-valued coupling strength between the dipole and the \(k^{\text{th}}\) mode of the bath and \(\textbf{X}_j\) is the \(j^{\text{th}}\) monomers position.  
Due to the presence of permanent dipoles, the multipolar gauge also introduces a non-zero self-dipole term:
\begin{multline}\label{Eq: SD}
    \hat{V}_{1} = \sum_k \frac{f_k^2}{\nu_k} \big( (\hat{\mathbf{d}}_1\cdot \textbf{e}_k)^2+(\hat{\mathbf{d}}_2\cdot \textbf{e}_k)^2\\ +2 \cos(\textbf{k}\cdot(\textbf{X}_1-\textbf{X}_2))(\hat{\mathbf{d}}_1\cdot \textbf{e}_k)(\hat{\mathbf{d}}_2\cdot \textbf{e}_k)\big) 
\end{multline}

Despite involving terms from the optical field, \(\hat{V}_{1}\) acts solely on the dimer. This term acts as a correction to the electrostatic interaction when in the multipolar gauge, and hence we absorb it into the electrostatic interaction with \(\hat{V} = \hat{V}_{0} + \hat{V}_{1}\) and \(Q_{ab}' = Q_{ab} + \langle a|\hat{V}_{1}|b \rangle\). 

The Hamiltonians presented in this subsection allow for the investigation of the effects of the permanent dipoles present in monomers on the optical properties of the dimer system. We will now further assume that the dimer state in which both monomers are excited can be disregarded as inaccessible because of its higher energy. As such, any coupling to this state is negligible for the eigenstructure generated. 

\subsection{Optical Environment}
Our model considers the dimer to be embedded in an unstructured, homogeneous medium, with the optical field treated as unpolarised and in a thermal equilibrium Gibbs state. This is defined through the density operator of the optical field as
\begin{align}
    \rho_B = \frac{\text{e}^{-\beta\hat{H}_O}}{Z_{O}},
\end{align}
where \( \beta \) is inversely proportional to the temperature \(\beta = 1/(k_b T) \) and \(Z_O =  \text{Tr}\{\exp{-\beta \hat{H}_O}\}\). The coupling between the dipoles of each monomer and the optical field is characterised by a continuous  spectral density \(J(\nu) = \sum_k |f_k|^2 \delta(\nu - \nu_k) \),
where \(\nu_k\) is strictly positive. Since we use the multipolar gauge in a homogeneous medium, the spectral density is cubic and contains a high-frequency cutoff \(\nu_c\). We define this spectral density as
\begin{align}\label{Eq: spect}
    J(\nu) = S\frac{\nu^3}{\nu_c^2}\text{e}^{-\frac{\nu}{\nu_c}},
\end{align}
where \(S = n \nu_c^2|\textbf{d}_\text{ref}|^2/8 \pi^2 \) is a dimensionless coupling constant. Here, \(n\) is the refractive index and the reference dipole \(|\textbf{d}_\text{ref}|\) is set to 1 Debye. Whenever the continuous definition is used, it is understood that all dipoles are transformed to the dimensionless vector \(\textbf{d}/|\textbf{d}_\text{ref}|\).  The transformation between the discrete and continuous definitions occurs via the transformation,
\begin{multline} \label{Eq: cont}
    \sum_k  f_k^2(\textbf{d}_{i}\cdot \textbf{e}_k)(\textbf{d}_{j}\cdot\textbf{e}_k)F(\nu_k)\\ 
    \rightarrow \frac{8\pi}{3} (\textbf{d}_{i} \cdot \textbf{d}_{j}) \int_0^\infty \text{d}\nu J(\nu)F(\nu),
\end{multline}
where \(F(\nu)\) is any energy-dependent function. We will use either the discrete or continuous definition of the optical field, selecting the form most appropriate for the context. To account for the reduced coupling to high-energy modes that do not couple as strongly to the molecular dipoles, while keeping the spectral density cubic for the most relevant modes, we set \(\nu_c = 10eV\). This also ensures that, for the system we are considering where the monomers are separated by at most a few nanometers, \(|\textbf{k}_c\cdot(\textbf{X}_1-\textbf{X}_2)| \ll 1\), meaning that we can extend the electric dipole approximation across the entire dimer, removing all position-dependent phase factors in the interaction Hamiltonian and the self-dipole term. Performing the integration in Eq.(\ref{Eq: cont}), we can now simplify Eq.(\ref{Eq: SD}) to
\begin{align}
    \hat{V}_1 = \lambda(\hat{\mathbf{d}}_1+\hat{\mathbf{d}}_2)^2,
\end{align}
where \(\lambda = \frac{16\pi \nu_c S}{3} \). Substituting in our values of \(\nu_c = 10eV\) and \(n = 1\), the dimensionless coupling constant is \(S = 1.29 \times 10^{-9}\) and \(\lambda = 1.08 n eV\). Due to the small value of $\lambda$ we neglect the contribution of the self-dipole energy in the main body of the paper.

\subsection{Weak Coupling Dynamics}

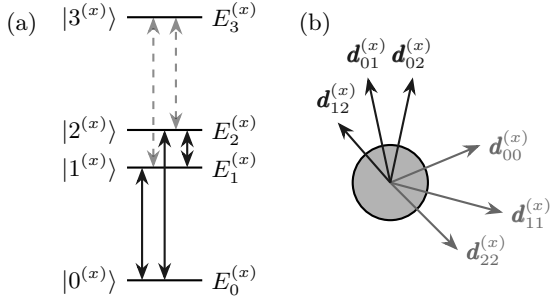
\begin{figure}[ht]
    
    \begin{minipage}{0.45\textwidth}
        \centering
        \begin{tikzpicture}[>=Stealth, thick, scale=1]

\def\Ezero{0}
\def\Eone{1.5}
\def\Etwo{2}
\def\Ethree{\Eone+\Etwo} 

\def\xleft{0}
\def\xmid{0}
\def\xright{0}

\draw[thick] (\xmid-0.5,\Ezero) -- (\xmid+0.5,\Ezero) node[right] {$E^{(x)}_0$};

\draw[thick] (\xleft-0.5,\Eone) -- (\xleft+0.5,\Eone) node[right] {$E^{(x)}_1$};

\draw[thick] (\xright-0.5,\Etwo) -- (\xright+0.5,\Etwo) node[right] {$E^{(x)}_2$};

\draw[thick] (\xmid-0.5,\Ethree) -- (\xmid+0.5,\Ethree) node[right] {$E^{(x)}_3$};

\node[left] at (\xmid-0.5,\Ezero) {$|0^{(x)}\rangle$};
\node[left] at (\xleft-0.5,\Eone) {$|1^{(x)}\rangle$};
\node[left] at (\xright-0.5,\Etwo) {$|2^{(x)}\rangle$};
\node[left] at (\xmid-0.5,\Ethree) {$|3^{(x)}\rangle$};

\draw[<->, thick,  black!90!white] (\xmid-0.3,\Ezero) -- (\xleft-0.3,\Eone);
\draw[<->, thick,  black!90!white] (\xmid+0,\Ezero) -- (\xright+0,\Etwo);
\draw[<->,dashed,opacity=0.5, thick,  black!90!white] (\xleft-0.15,\Eone) -- (\xmid-0.15,\Ethree);
\draw[<->,dashed,opacity=0.5, thick,  black!90!white] (\xright+0.15,\Etwo) -- (\xmid+0.15,\Ethree);
\draw[<->, thick, black!90!white] (\xmid+0.3,\Eone) -- (\xleft+0.3,\Etwo);
\node [label={[shift={(-1.9,3.0)}](a)}] {};     
            \def\shift{3}
\def\shiftup{1.3}

\draw[fill=black!30!white] (\shift,\shiftup) circle (0.5);

\draw[->, white!40!black, thick] (\shift+0,\shiftup+0) -- (\shift+1.2,\shiftup+0.5) node[right] {\(\pmb{d}^{(x)}_{00}\)};
\draw[->, white!40!black, thick] (\shift+0,\shiftup+0) -- (\shift+1.5,\shiftup-0.4) node[right] {\(\pmb{d}^{(x)}_{11}\)};
\draw[->, white!40!black, thick] (\shift+0,\shiftup+0) -- (\shift+0.9,\shiftup-0.9) node[right] {\(\pmb{d}^{(x)}_{22}\)};
\draw[->, black!90!white, thick] (\shift+0,\shiftup+0) -- (\shift-0.3,\shiftup+1.4) node[above] {\(\pmb{d}^{(x)}_{01}\)};
\draw[->, white!10!black, thick] (\shift+0,\shiftup+0) -- (\shift+0.3,\shiftup+1.4) node[above] {\(\pmb{d}^{(x)}_{02}\)};
\draw[->, black!90!white, thick] (\shift+0,\shiftup+0) -- (\shift-0.7,\shiftup+0.8) node[above] {\(\pmb{d}^{(x)}_{12}\)};
\node [label={[shift={(2.0,3.0)}](b)}] {};
\end{tikzpicture}
        
    \end{minipage}
    
    \caption{\justifying (a) Energy level structure for the system in it's eigenbasis. The dashed lines going to the third excited state show that these transitions will be neglected. (b) Diagram of the dimers dipoles in its eigenbasis after the electric dipole approximation has been applied across the dimer. The darker arrows represent the transition dipoles of the dimer while the lighter ones show the permanent dipoles.}
    \label{fig2}
\end{figure}

We now use the secularised Bloch-Redfield master equation (SBRME) to calculate the transition rates of the dimer system which requires the system Hamiltonian to be diagonalised. Here, any states and scalar values in the dimer eigenbasis of a system \((x)\) are denoted by \(\ket{a^{(x)}}\) and \(S^{(x)}\), respectively. Restricting our analysis to the single excitation manifold and ground state, the system Hamiltonian can be written as \(H^{(x)}_S = \sum_{a=0}^2 E^{(x)}_{a}\ket{a^{(x)}}\bra{a^{(x)}}\). The difference between eigenenergies is denoted as \(\omega^{(x)}_{ab} = E^{(x)}_{a}-E^{(x)}_{b}\). A depiction of the energy level structure is shown in  Fig(\ref{fig2}a). The interaction Hamiltonian transformed to the diagonal dimer frame takes the general form of
\begin{align}
    \tilde{H}_I = \sum_k \sum_{a,b=0}^3 -if_k \ |a^{(x)}\rangle \langle b^{(x)}| (\textbf{d}_{ab}^{(x)}\cdot \mathbf{e}_k) \hat{a}_k^\dagger +\text{h.c},
\end{align}
where \(\textbf{d}^{(x)}_{ab} = \bra{a^{(x)}}U^{(x)} (\mathbf{\hat{d}}_1+\mathbf{\hat{d}}_2)U^{(x)\dagger} |b^{(x)}\rangle\) and \(U^{(x)}\) is the unitary operator that takes the system from the site basis to the eigenbasis. \text{h.c} represents the Hermitian conjugate. An example of the dipoles in the eigenbasis is shown in Fig(\ref{fig2}b).

The time evolution of the population for any eigenstate \(\rho_{aa}\), according to the SBRME, is
\begin{multline}\label{Eq: Redfield}
     \frac {\partial }{\partial t}\rho^{(x)}_{aa}(t)= \sum_{b} |\textbf{d}^{(x)}_{ab}|^2 \big[\gamma(\omega_{ab}^{(x)})\rho^{(x)}_{bb}(t)\\
     - \gamma(-\omega_{ab}^{(x)})\rho^{(x)}_{aa}(t)\big],
\end{multline}  
where \(\rho_{ab}^{(x)} = \ket{a^{(x)}}\bra{b^{(x)}}\). The rate term \(\gamma\) is defined as
\begin{align}
    \gamma(\omega) = \Bigg{\{} \begin{matrix}
        \frac{8\pi}{3}J(\omega)(N(\omega)+1) & \text{if } \omega >0,\\
         \frac{8\pi}{3}J(|\omega|)N(|\omega|) & \text{if } \omega <0,
    \end{matrix}
\end{align}
where \(N(\omega)\) is the Bose-Einstein distribution expectation value for the occupations of modes,  equal to \( (\exp(\beta \omega)-1)^{-1}\), ensuring that the system maintains detailed balance. When \(\omega\) is positive, \(\gamma\) depends on the rate of spontaneous and stimulated emission of a photon, while negative values of \(\omega\) correspond to the rate of photon absorption. Here we can see that, according to the SBRME, the rate of population transfer between two states is directly proportional to the square of the corresponding transition dipole of the dimer and does not depend on the permanent dipoles of the dimer states. In the appendix, we use the polaron transformation to see the effects of the permanent dipoles that are not captured by the SBRME but no significant changes are found. We will use these equations to calculate the transition rates of different dimer configurations and identify the conditions for dark state formation.

\section{Dark States}\label{Sec: Rates}
Dark states occur when the transition dipole between an excited and ground state is very small. This may occur when symmetries in molecular orbitals cause a transition to be naturally forbidden, or when superpositions of excited states cause the composite dipoles contributing to the dimers' transition dipole to destructively interfere. Permanent dipoles in the monomers can significantly affect the electrostatic coupling between the monomers and create a different eigenstructure of the dimer. This, in turn, changes the nature of dipole interference.

In this section, we explore how permanent dipoles can affect the optical properties of dimers that contain dark states. To keep the analytics of the transition rates tractable, we focus on cases that are analytically diagonalisable. Each subsection focuses on a different eigenstructure where dark states can form.

\subsection{Direct Exciton Coupling}

The situation most studied for dimers in the literature occurs when the only electrostatic interaction between the monomers occurs between the transition dipoles. The resulting eigenbasis causes these transition dipoles to interfere with each other, creating a dark and bright state. In this section, we investigate how permanent dipoles affect the properties of dimers when they do not induce any coupling between the ground and excited states of the monomers. These systems do not rely on permanent dipoles to form. The monomer frame Hamiltonian is of the form
\begin{align}
    H_{S}^{(A)} =  \sum_{a=0}^2 \epsilon_a\ket{a}\bra{a} +(Q_{12} \ket{1}\bra{2}+h.c.),
\end{align}
where we have defined \(\epsilon_a = E_a+Q_{aa}- E_0 - Q_{00}\). Assuming that \(Q_{12}\) is real, the energy levels of the dimer are \(E^{(A)}_{1} = \frac{1}{2}(\epsilon_1+\epsilon_2 - \omega^{(A)}_{12})\) and \(E^{(A)}_{2} = \frac{1}{2}(\epsilon_1+\epsilon_2 + \omega^{(A)}_{12})\) where \(\omega^{(A)}_{12}=  \sqrt{(\epsilon_1-\epsilon_2)^2+4 Q_{12}^2}\). The ground state of the dimer and its energy remain unchanged from the site basis, while the excited states of the dimer become delocalised with the asymmetric transformation:
\begin{subequations}\label{Eq: sym basis}
\begin{align}
    &\ket{1^{(A)}} = \cos \frac{\chi^{(A)}}{2}\ket{1} + \sin \frac{\chi^{(A)}}{2}\ket{2},\\
    &\ket{2^{(A)}} = \cos \frac{\chi^{(A)}}{2} \ket{2}- \sin \frac{\chi^{(A)}}{2}\ket{1},
\end{align}
\end{subequations}
where, \(\sin(\chi^{(A)}) = \frac{2Q_{12}}{\omega_{12}}\) and \(\cos(\chi^{(A)}) = \frac{\epsilon_1-\epsilon_2}{\omega_{12}}\). For symmetric configurations of a homodimer, it is expected that the energy resulting from the permanent dipole electrostatic interaction is the same for each monomer. In these cases, \(\epsilon_1 = \epsilon_2\) and both excited states of the dimer are perfectly delocalised across each monomer (${\chi^{(A)}}=\pi/2$). 

The transition rates between the ground and excited states are proportional to the square of the corresponding dimer transition dipole,
\begin{align}
    \textbf{d}^{(A)}_{01} = \cos(\frac{\chi^{(A)}}{2}) \pmb{\mu}_1 + \sin (\frac{\chi^{(A)}}{2}) \pmb{\mu}_2,\\
    \textbf{d}^{(A)}_{02} = \cos(\frac{\chi^{(A)}}{2}) \pmb{\mu}_2 - \sin (\frac{\chi^{(A)}}{2}) \pmb{\mu}_1.
\end{align}
Just as when no permanent dipoles are involved, the transition between any excited and ground state can only be forbidden in cases where \(\pmb{\mu}_1\) and \(\pmb{\mu}_2\) are either perfectly aligned or anti-aligned. The excited state on which the dark state forms depends on this and the sign of \(Q_{12}\), with a bright state forming on the other. For these transitions in the weak-coupling regime, the permanent dipoles affect only the energy levels and the localisation of the excited states. 

For this eigenstructure, permanent dipoles have their most pronounced impact on the transition rates between excited states. In their absence, such transitions generally require external mechanisms, such as vibrational coupling. However, the presence of permanent dipoles enables these transitions to occur optically. The rate at which the first excited state becomes occupied from the second excited state is
\begin{align}
    \frac{d(\rho_{22}^{(A)}\rightarrow\rho_{11}^{(A)})}{dt}= \gamma(\omega_{21}) \frac{|Q_{12}|^2}{2 \omega_{12}^2} |\pmb{\Delta}_1-\pmb{\Delta}_2|^2\rho^{(A)}_{22}(t). 
\end{align}
Here, the changes in permanent dipoles of the monomers become transition dipoles between the excited states of the dimer. These behave in a manner similar to that in which localised vibrations can induce transitions between excited states. However, there are two key differences.

The first is that, unlike localised vibrations, the correlation between the interactions that act on the excited states of the monomers depends on the relative orientation of \(\pmb{\Delta}_1\) and \(\pmb{\Delta}_2\). When these dipoles are orthogonal to each other, there is no correlation between the two, as with vibrations, leading to a summation of the two difference dipoles. For non-orthogonal dipoles, they interfere with each other to increase or decrease the transfer rate between the excited states. When the change in permanent dipoles is the same for both monomers, the permanent dipole interactions are perfectly correlated for each monomer, and the transitions between excited states no longer occur. When the dipoles are anti-aligned, we attain a rate that is the square of the sum, leading to a factor of 4 enhancement (twice that of the uncorrelated rate).

\begin{figure}
    \centering
    \includegraphics[width=\linewidth]{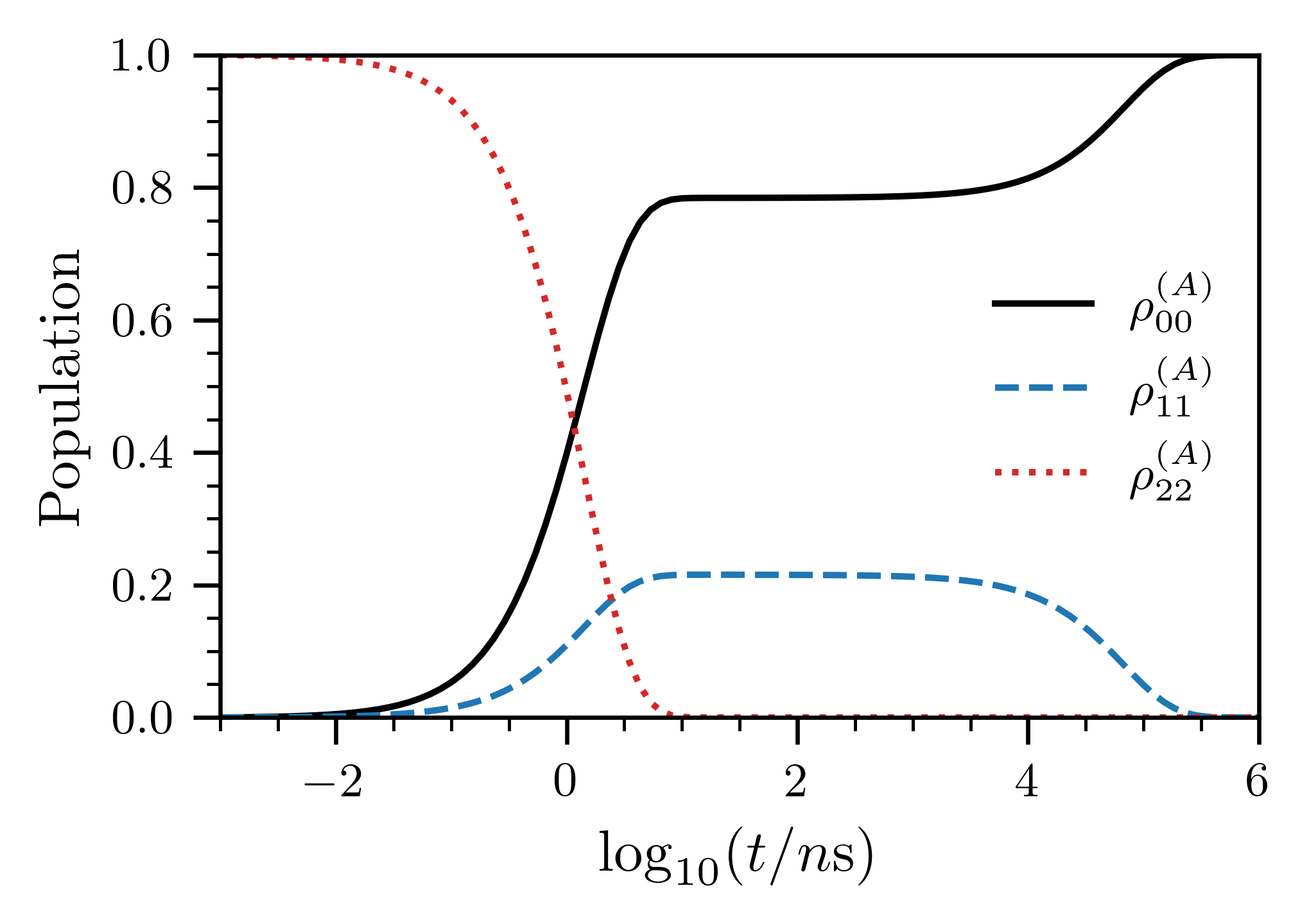}
    \caption{\justifying Population dynamics of system (A) beginning in state \(\ket{2^{(A)}}\) for logarithmic time. The energy of the eigenstates are \(\omega_{02}^{(A)} = 2.8\)eV and \(\omega_{01}^{(A)} = 2.5\)eV. The transition dipoles of both monomers are \(|\pmb{\mu}_j|\) = 10 D and they are parallel to each other. The permanent dipoles are such that \(|\pmb{\Delta}_1-\pmb{\Delta}_2|\) = 60 D. Since \(\ket{1^{(A)}}\) is a perfect dark state, all decays from \(\ket{1^{(A)}}\) occur via an excitation to \(\ket{2^{(A)}}\). The rate of this depends on the temperature of the optical field which is set to 300K.}
    
    \label{fig: Pop}
\end{figure}

The second key difference is that, while vibrational interactions can couple to a bath with a temperature different from that of the transition dipoles, this is not true for the permanent dipoles. This means that the dimer will always maintain detailed balance, and therefore, this mechanism cannot be used to improve the power output of the dimer in quite the same way as vibrations. As can be seen in Fig.(\ref{fig: Pop}), the permanent dipoles of this system can increase the time for which an exciton is present, but the dimer will still return to equilibrium.

\subsection{Indirect Coupling}
The permanent dipole of the ground state of one monomer can interact with the transition dipole of the other through the electrostatic interaction defined in Eq.(\ref{Eq: Coulomb Gen}). This can cause the single-monomer excitation states to become coupled to the zero-excitation states, resulting in an indirect coupling of the single excitation states.

\begin{figure}
    \centering
    \includegraphics[width=\linewidth]{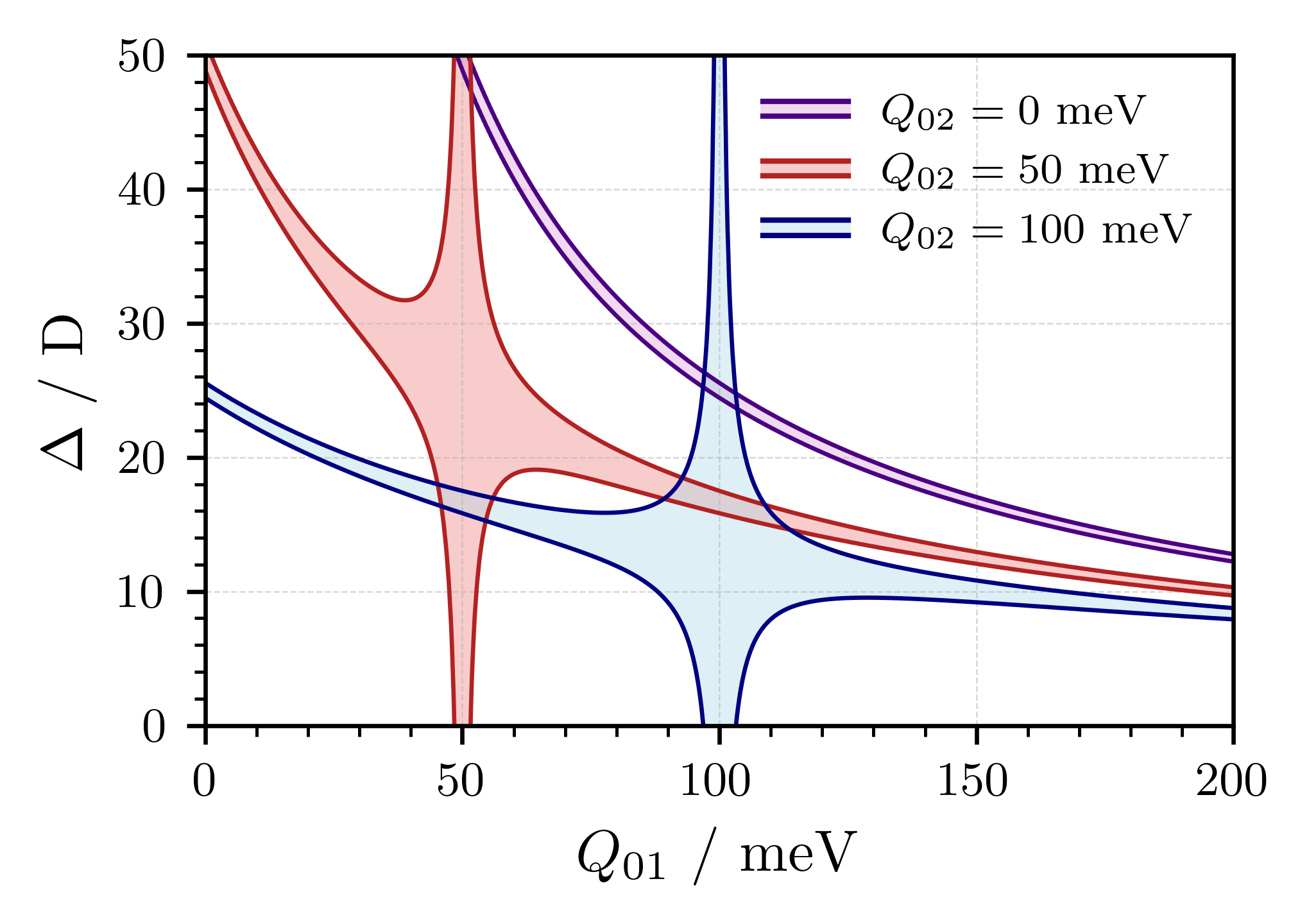}
    \caption{\justifying Conditions for dark states to form on state \(\ket{2^{(B)}}\) for anti-parallel alignment of monomers. The shaded regions indicate the configurations where \(|\textbf{d}_{02}^{(B)}|^2 < |\pmb{\mu}_j|^2 \times 10^6\). The purple, red and blue colours represent the dimers whose \(Q_{02}\) is 0, 50 and 100 meV respectively. The two monomers have energies of \(\epsilon_j' = \) 2.5 eV and anti-parallel transition dipoles with a strength of 2 D. The change in permanent dipoles is equal and opposite for both monomers, and their magnitudes are given on the y-axis.}
    
    \label{fig: Parallel}
\end{figure}

In this subsection, we investigate the case where the only coupling between states in the site basis is between the ground and single excitation states. This allows us to isolate the resulting effects. To ensure that the Hamiltonian is analytically diagonalisable, we assume that \(\epsilon = \epsilon_1 = \epsilon_2\). The Hamiltonian is subsequently of the form
\begin{align}
    H_{S}^{(B)}=  \sum_{j=1}^2 \epsilon\ket{j}\bra{j}+(Q_{0j} \ket{0}\bra{j}+h.c.),
\end{align}
where we have removed an identity term. This results in the energy of the first excited state remaining unchanged while the ground and second excited states of the dimer become
\begin{align}
    E^{(B)}_0 &= \frac{ \epsilon-\omega_{02}^{(B)}}{2},\\
   E^{(B)}_2 &= \frac{ \epsilon+\omega_{02}^{(B)}}{2},
\end{align}
where 
\begin{equation}
    \omega_{02}^{(B)} =\sqrt{4(|Q_{01}|^2+|Q_{02}|^2)+\epsilon_1^2} .
\end{equation}
For this system, it is always the case that \(E^{(B)}_0<E^{(B)}_1<E^{(B)}_2\) where the eigenstates of the system are as follows:
\begin{align}
    |0^{(B)}\rangle &= \cos{\Big(\frac{\chi^{(B)}}{2}\Big)}\ket{0} -\sin{\Big(\frac{\chi^{(B)}}{2}\Big)}\big(D_2
    |1\rangle +D_1|2\rangle\big),\\
    |1^{(B)}\rangle &= D_1|1\rangle - D_2|2\rangle,\\
    |2^{(B)}\rangle &= \cos{\Big(\frac{\chi^{(B)}}{2}\Big)}\big(D_2
    |1\rangle +D_1|2\rangle\big) + \sin{\Big(\frac{\chi^{(B)}}{2}\Big)}\ket{0}.
\end{align}
Here, \(D_j = \frac{Q_{0j}}{\sqrt{|Q_{01}|^2+|Q_{02}|^2}}\) is the normalized driving term, while \(\chi^{(B)}\) is defined through the equations \(\sin(\chi^{(B)}) = 2\frac{\sqrt{|Q_{01}|^2+|Q_{02}|^2}}{\omega_{02}^{(B)}}\) and \(\cos(\chi^{(B)}) = \frac{\epsilon}{\omega_{02}^{(B)}}\). The degree to which the states are localised depends on the relative strength of the coupling between the ground and excited states of each monomer. Unlike the direct coupling case, the ground state of the dimer now includes the excited states of the monomers. This results in the dimers' transition dipole between the ground and excited states containing contributions from the permanent dipoles of the monomers. For the first excited state, this is
\begin{multline}
    \textbf{d}^{(B)}_{01} = \cos(\frac{\chi^{(B)}}{2}) (D_1\pmb{\mu}_1- D_2\pmb{\mu}_2)\\
    -\sin(\frac{\chi^{(B)}}{2})D_1D_2(\pmb{\Delta}_1- \pmb{\Delta}_2).
\end{multline}
There are many ways in which this state can become dark, but the simplest occurs in the highly symmetric case where the magnitude of the driving forces and respective dipoles are the same for both monomers. When the change in permanent dipole for each monomer is aligned, their permanent dipoles do not directly contribute to the transition dipole. A dark state can then form if either the transition dipoles of each monomer are aligned and the driving they experience is equal, or if they are anti-aligned and \(Q_{01}+Q_{02} =0\). In cases where the change in permanent dipole is not the same for both monomers, the permanent dipoles can interfere with the transition dipoles. This can allow dark states to form even in situations where the monomer transition dipoles are orthogonal to each other. 
The transition dipole between the ground and the second excited state of the dimer is
\begin{multline}
    \textbf{d}^{(B)}_{02} = \cos(\chi^{(B)})(D_2\pmb{\mu}_1+ D_1\pmb{\mu}_2)\\
    -\frac{\sin(\chi^{(B)}) }{2}(D_2^2\pmb{\Delta}_1+ D_1^2\pmb{\Delta}_2).
\end{multline}
Unlike \( \textbf{d}^{(B)}_{01}\), here the relative contribution of the permanent dipoles of each monomer is not necessarily equal. This can allow the effect of permanent dipoles of the monomers on the transition rate of the dimer to be tuned by varying the electrostatic driving energy, a property dependent on the relative orientation and separation of the two monomers. As can be seen in Fig(\ref{fig: Parallel}), whenever all the constituent dipoles of the dimer are parallel, there are many ways in which dark states can form. The vertical shaded regions show the simple symmetric case where a dark state can form because each dipole destructively interferes with its counterpart on the other monomer. The horizontal areas show the regions where the change in permanent dipoles destructively interfere with the transition dipoles. This is of significant interest, as when both effects intersect, the region of the configuration space where the state is considered to be dark is much larger. This suggests that the combination of both interferences may make the dark states more robust, a theory which we investigate more in Sec.(\ref{Sec: Robust}). 
\\

The purple line in Fig.(\ref{fig: Parallel}) represents an excited state that is entirely localised on a single monomer. We find that dark states can form in this region when 
\begin{align}\label{Eq: In Dark Con}
     \epsilon \pmb{\mu}_1 =Q_{01} \pmb{\Delta}_1,
\end{align}
and \(Q_{02} = 0\). Such a configuration would also result in \(\ket{1^{(B)}} = \ket{1}\). However, this does not mean that the dimer would act as two separate monomers. The transition rates between the two excited states are proportional to the square of the dimer transition dipole
\begin{multline}\label{Eq: In tran}
    \textbf{d}^{(B)}_{12} = \sin(\frac{\chi^{(B)}}{2}) (D_1\pmb{\mu}_1- D_2\pmb{\mu}_2) \\+\cos(\frac{\chi^{(B)}}{2})D_1D_2(\pmb{\Delta}_1-\pmb{\Delta}_2).
\end{multline}
Using equations (\ref{Eq: In Dark Con}) and (\ref{Eq: In tran}), it can be seen that for the state \(|2^{(B)}\rangle\) to be a localised dark state, it must become optically accessible from the state \(|1^{(B)}\rangle\). These properties of localised dark states remain true when \(\epsilon_1\) or \(\epsilon_2\) are not equal to each other, allowing for localised dark state to exist at the lower-energy excited state of the dimer.

However, it is important to note that the creation of a localised dark state in this form relies on multipolar electrostatic interactions. When the point-dipole approximation is applied for calculating the electrostatic terms in Eq.(\ref{Eq: Coulomb Gen}), the electric field that a monomer experiences from the other is the same for each of its dipoles. This means that when the transition and permanent dipoles are aligned, any driving force on the transition dipole must also affect the permanent dipoles. By calculating the eigenbasis transition dipole for a single monomer system \((S)\) that is subject to a uniform electric field \(\pmb{E}\), we find that
\begin{align}\label{Eq: Uniform pert}
    \textbf{d}^{(S)}_{01}=\frac{(\epsilon +\pmb{\Delta}_1\cdot \pmb{E}) \pmb{\mu}_1-(\pmb{\mu}_1\cdot \pmb{E})\pmb{\Delta}_1}{\sqrt{(\frac{\epsilon+\pmb{\Delta}_1\cdot \pmb{E}}{2})^2+(\pmb{\mu}_1\cdot \pmb{E})^2}}.
\end{align}
Note that in the numerator, any components of \(\pmb{\Delta}_1\) that are parallel to \(\pmb{\mu}_1\) vanish. This means that while permanent dipoles can contribute to the transition rate of a monomer in a uniform field, they cannot effectively interfere with the original transition dipole. As a result, localised bright and dark states caused by interference between the monomer transition and permanent dipoles must rely on the presence of multipolar electrostatic interactions that are more present when monomers are in close proximity to each other. Focusing on the denominator in Eq.(\ref{Eq: Uniform pert}), we see that the eigenstate transition dipole can still be suppressed or enhanced by the monomers coupling to a uniform electric field. However, an arbitrary amount of suppression cannot be achieved, as, while in the high coupling limit \(\pmb{E}\) is inversely proportional to the eigenstate transition dipole, it is also proportional to the energy splitting \(\omega^{(S)}_1\). Since the transition rate is proportional to \((\omega^{(S)}_1)^3 |\textbf{d}^{(S)}|^2\), this limit would result in an increased decay rate from the excited state.  Note that reaching this limit requires very strong coupling, which would likely break other assumptions that have been made in the model. 

The dependency on multipolar electrostatic interactions for the permanent dipoles to interfere with the transition dipoles does necessarily apply to delocalised site basis excited states because, then the electric field experienced by the monomer can become entangled with the state of the dimer.

\subsection{Mixed Coupling}
Finally, we explore the optical properties of dimers when both direct- and indirect-coupling are in play. Keeping the eigenbasis analytic requires a high degree of symmetry in the Hamiltonian, so we shall restrict any analysis in this section to that of homodimers, where any electrostatic interactions are symmetric. The Hamiltonian of the dimer that we shall consider in the site basis is 
\begin{multline}\label{Eq: Inter Ham}
    H_{S}^{(C)}= 
    \epsilon (\ket{1}\bra{1}+\ket{2}\bra{2})\\
    +(Q_{G} (\ket{0}\bra{1}+\ket{0}\bra{2}) +Q_{X}\ket{1}\bra{2} +h.c.),
\end{multline}
The coefficient \(Q_{G}\) comes from the electrostatic driving energy between the ground and excited states and is equal to \(Q_{01}\) and \(Q_{02}\). We also define \(Q_{X} = Q_{12}\) to keep the notation of this subsection consistent. Such a Hamiltonian can form naturally when the point-dipole approximation is used to calculate the electrostatic energy in situations when the dipoles of each monomer are aligned. The eigenbasis of this system is
\begin{align}
    |0^{(C)}\rangle &= \cos{\Big(\frac{\chi^{(C)}}{2}\Big)}\ket{0} -\sin{\Big(\frac{\chi^{(C)}}{2}\Big)}\frac{|1\rangle + |2\rangle}{\sqrt{2}},\\
    |1^{(C)}\rangle &= \frac{|1\rangle - |2\rangle}{\sqrt{2}},\\
    |2^{(C)}\rangle &= \cos{\Big(\frac{\chi^{(C)}}{2}\Big)}\frac{|1\rangle + |2\rangle}{\sqrt{2}} + \sin{\Big(\frac{\chi^{(C)}}{2}\Big)}\ket{0},
\end{align}
where \(\cos(\chi^{(C)}) = \frac{\epsilon+Q_X}{\omega_{02}^{(C)}}\), \(\sin(\chi^{(C)}) = \frac{2\sqrt{2} Q_G}{\omega_{02}}\) and \(\omega^{(C)}_{02} = \sqrt{8 Q_{01}^2+ (\epsilon_1+ Q_{12})^2}\). Here, a higher value of \(\chi^{(C)}\) can be loosely associated with a relative increase in intermediate coupling through the ground state. Although we will restrict our analysis to configurations where \(\pmb{\mu}_1\), \(\pmb{\mu}_2\), \(\pmb{\Delta}_1\) and \(\pmb{\Delta}_2\) are aligned, we make no such restriction when presenting any of the equations. The transition dipole between the  \(\ket{\tilde{1}}\) and ground state is 
\begin{align}
    &\textbf{d}^{(C)}_{01} = \cos(\frac{\chi^{(C)}}{2})(\pmb{\mu}_1- \pmb{\mu}_2) -\sin(\frac{\chi^{(C)} }{2})(\pmb{\Delta}_1-\pmb{\Delta}_2).
\end{align}
The symmetry we have imposed for our analysis ensures that this state is dark independent of the value of \(\chi^{(C)}\) due to the destructive interference between the dipoles of each monomer. Since the eigenbasis transition dipole between the two excited states is
\begin{align}\label{Eq: mixed inter excited tran}
    &\textbf{d}^{(C)}_{12} = \sin(\frac{\chi^{(C)}}{2})(\pmb{\mu}_1- \pmb{\mu}_2) +\cos(\frac{\chi^{(C)} }{2})(\pmb{\Delta}_1-\pmb{\Delta}_2),
\end{align}
decay paths through the \(\ket{2^{(C)}}\) state will be forbidden even if \(\epsilon_1 > \epsilon_2 \). The transition dipole between the ground state and \(\ket{\tilde{2}}\) can be given by the following equation:
\begin{align}
    &\textbf{d}^{(C)}_{02} = -\cos(\chi^{(C)})(\pmb{\mu}_1+ \pmb{\mu}_2) +\sin(\chi^{(C)} )(\pmb{\Delta}_1+ \pmb{\Delta}_2),
\end{align}
where here, each monomer's dipoles constructively interfere with each other. However, it is still possible for \(|2^{(C)}\rangle\) to become dark if the permanent and transition dipoles constructively interfere. This occurs when \((\epsilon+Q_X)(\pmb{\mu}_1+\pmb{\mu}_2) = 2 \sqrt{2} Q_G(\pmb{\Delta}_1+\pmb{\Delta}_2) \) and can cause every optical transition in the zero and single excitation manifold to no longer occur. This is not possible with out permanent dipoles, as otherwise, the total dipole magnitude is conserved when the monomer couple. Here, however, the transition and permanent dipole can mix with each other, effectively moving dipole strength to dimer permanent dipoles. Excitations in such states would have no way to emit a photon, and so may be particularly useful as ways of storing energy for longer times. These states may therefore only be accessed through other mechanisms such as vibrational effects or by eigenstructure changes in the presence of external electric fields.

In order to investigate such a system outside of a perfectly symmetric Hamiltonian, in the following section, we use numerical calculations to study disorder.

\section{Robustness of Dark States}\label{Sec: Robust}
We have shown that the permanent dipoles of monomers can affect the optical properties of the dimer and, under certain circumstances, these permanent dipoles can cause dimers to have dark states through indirect coupling between the excited states of the monomers. However, since these systems are difficult to investigate analytically, it is not immediately apparent how they will behave outside of their analytic regime. In real-world situations, the dimer system would be subject to thermal fluctuations and vibrations that perturb the states of each monomer, moving the Hamiltonian of the dimer outside of analytic regimes. In this section, we numerically investigate how robust the dark states of system (C) are to fluctuations in their energy levels, analogous to those caused by thermal fluctuations.

\begin{figure}
    \centering
    \includegraphics[width=\linewidth]{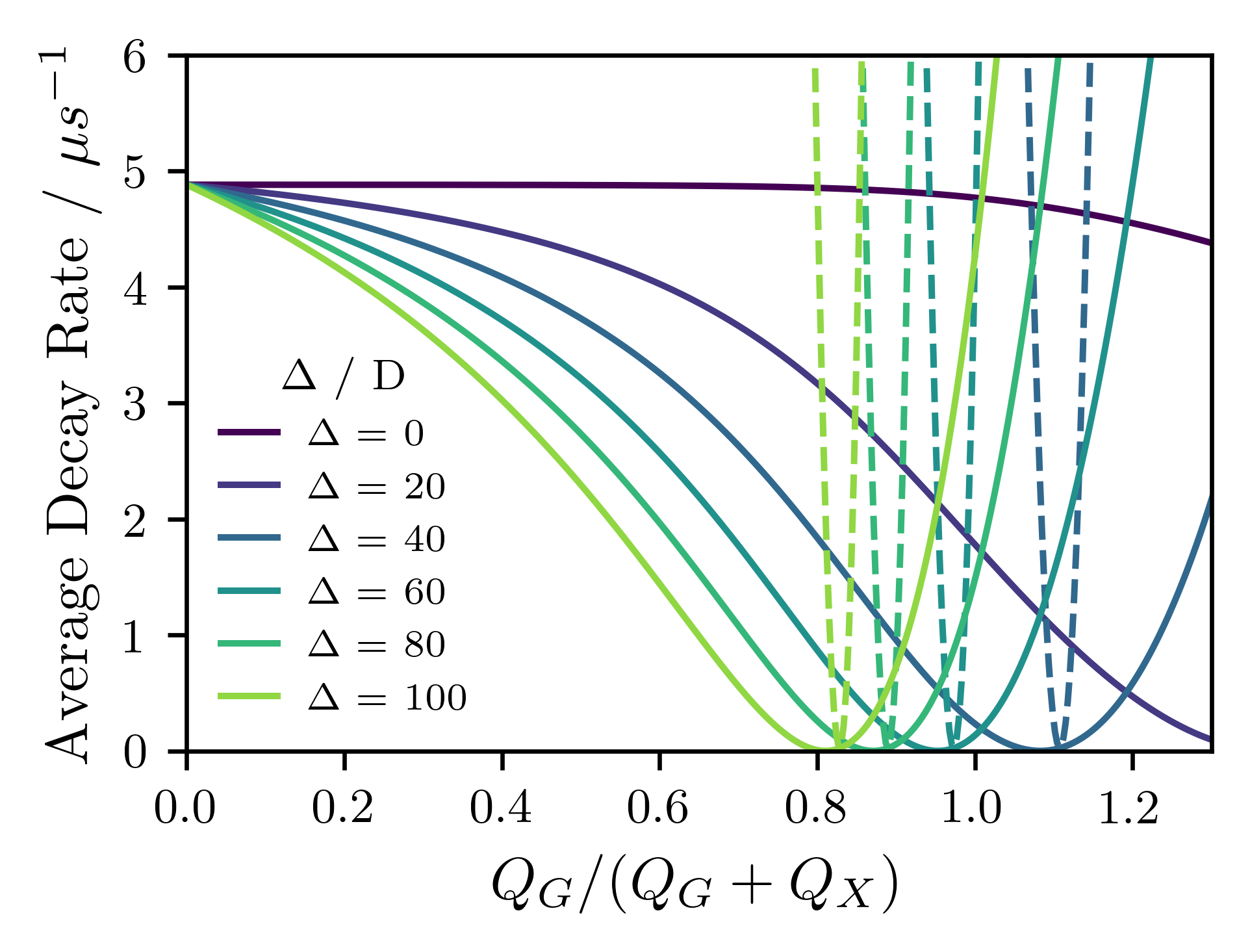}
    \caption{\justifying Average decay rate of systems (C) when the monomers energy levels are subjected to uncorrelated random Gaussian perturbations of \(25\) meV. The solid lines show the rate of decay from the \(\ket{1^{(C)}}\) state while the dashed lines shows the rate from \(\ket{2^{(C)}}\). The parameters used in this system are \(\epsilon_1 = \epsilon_2 = 2.4\)eV and \(|\pmb{\mu}_j| = 10\) D. \(Q_G\) and \(Q_X\) are such that the unperturbed energy splitting \(\omega_{12}^{(C)} = 150\) meV and \(Q_G\) is always positive. Each perturbation consists of the effective energy level for each monomer and each symmetric coupling term to be adjusted according to a Gaussian distribution with \(\mu = 0\) and \(\sigma = 25\) meV.  The data was calculated by applying 1000 random perturbations to each system and calculating the transition rate in the resulting eigenbasis.}
    \label{fig: state 1 rob}
\end{figure}

Since vibrations typically act locally on each monomer and at time scales much slower than the relaxation time of the electrons, we model their effects by randomly perturbing each monomer's energy level according to a Gaussian distribution with a standard deviation of \(\sigma = 25\) meV (\(k_B T\) at room temperature). To compare the robustness of direct and indirect coupling, we use the mixed coupling Hamiltonian defined in Eq.(\ref{Eq: Inter Ham}). In this section, the unperturbed monomer energies \(\epsilon_1 = \epsilon_2 = 2.4\)eV and the coupling terms \(Q_X\) and \(Q_G\) are such that the energy splitting between the two excited states is fixed at 150 meV, the dark state being the lower energy of the two. This avoids any potential issues that may arise from the secular approximation. Before any perturbations are added to the system, each solid line, representing the average transition rate from \(\ket{1^{(C)}}\) to the ground state,  in Fig.\ref{fig: state 1 rob} would have a transition rate of zero. The graph shows the average transition rates for 1000 different fluctuations in the site energy distributed as described above.
When \(Q_G = 0\), the excited states of the monomers are completely coupled to each other directly, depicted by the dark purple line showing the relative robustness of a dark state where no permanent dipoles are involved. As we move along the x-axis, the indirect coupling term \(Q_G\) becomes larger comparatively until it is the sole contributor at \(Q_G/(|Q_G|+|Q_X||) = 1\). Past this point, \(Q_X\) becomes negative. Fig(\ref{fig: state 1 rob}) shows that as the combination of direct and indirect coupling is tuned to the change in the permanent dipole, the dark state of \(\ket{1^{(C)}}\) can be significantly more robust to site energy fluctuations. The increase in robustness stems from the permanent dipoles of the monomers destructively interfering with their transition dipoles, meaning the state may remain dark even when the energy fluctuations cause the exciton to become more localised. The minimum average decay rate for each dipole strength is analogous to the configurations found at the centre of the shaded regions in Fig. (\ref{fig: Parallel}). Lesser values of \(Q_X\) can result in the most robust configurations having a smaller change in permanent dipole. The dashed lines show the relative rate of transition from \(\ket{2^{(C)}}\) to the ground state. For these states, the dipoles of each monomer constructively interfere with each other. This state can also become dark when the transition dipoles destructively interfere with the permanent dipoles. This occurs when \((\epsilon+Q_X)(\pmb{\mu}_1+\pmb{\mu}_2) = 2 \sqrt{2} Q_G(\pmb{\Delta}_1+\pmb{\Delta}_2) \). Since the transition dipoles of each monomer constructively interfere with each other for the \(\ket{2^{(C)}}\) to ground-state transition, any deviation in the relative contribution of the transition and permanent dipoles has a significantly larger effect. This causes a much steeper dip in the average transition rate when varying the coupling type ratios compared to \(\ket{2^{(C)}}\). when \(Q_X = 0\), the decay rate from \(\ket{2^{(C)}}\) is 100 times faster than \(\ket{1^{(C)}}\) after averaging over the fluctuations. The transition rate between the excited states at every point in the figure is so small that it can be ignored. This does not account for fluctuations in the dipole size, orientation, or the coupling between monomers.

\section{Conclusion }\label{Sec: Conc}
Dark states in dimers hold significant potential to improve the efficiency of light-harvesting and energy transport systems. However, these dark states must be tailored so that they are suitable for their role. The role of permanent dipoles in these processes has often been overlooked. By accurately accounting for the effects of permanent dipoles, we demonstrated that optical transfer between excited states can become enabled even when dark states form in the typical manner. Moreover, permanent dipoles facilitate indirect excitonic coupling between monomers via the ground state, leading to interference between the permanent and transition dipoles. This can allow for dark states in homodimers to be more localised, potentially allowing for easier transportation of excitons.

When both direct and indirect couplings are utilised in tandem, the robustness of dark states to energetic perturbations can be significantly enhanced, provided the dimer system is appropriately calibrated. This increased robustness can reduce the rate of recombination in noisy environments, improving the efficiency of light-harvesting and energy-transport systems.  

Overall, we have shown that when monomers are in close proximity to each other, permanent dipoles can have a significant effect on the emergent optical behaviour. This motivates further study into larger systems with a more rigorous modelling of environmental effects and the impact of strong coupling to cavity modes, enhancing the strength of permanent dipole effects. 
\section*{Acknowledgements}
The work of M. Freed was supported by funding from the University of Surrey. 
A. Burgess thanks the Leverhulme Trust for support through grant number RPG-2022-335.

\appendix*

\section{Effects of Permanent Dipoles on the Optical Master Equation}\label{App: Polaron}
Permanent dipoles in the constituent monomers often lead to permanent dipoles being present in the dimer system. Since the SBRME equation only includes terms up to second order in dipole coupling, any affect of the dimers' permanent dipoles are not accounted for. To remedy this, in this section we make use of the optical polaron transformation to include the fourth-order dipole-coupling terms that involve the permanent dipoles in the transition rates~\cite{burgess_strong_2023}. The polaron transformation incorporates the dimer's permanent dipole displacement of the optical modes into the definition of the state, "dressing" the system with a cloud of virtual photons. For our model, this transformation is generated by the unitary operator, \(P = B(\mathcal{Z}^{(x)})\)
where 
\begin{align}\label{Eq: B}
    B(\alpha) = \exp \big(i\sum_k\frac{1}{\nu_\textbf{k}} (\alpha_k \cdot\textbf{e}_k)a^\dagger_k +(\alpha_k^*\cdot\textbf{e}^*_k) a_k\big), 
\end{align}
and \(\mathcal{Z}\) is the dimer permanent dipole operator defined as
\begin{align}
    \mathcal{Z}^{(x)}_k = f_k\sum_{a=0}^3\pmb{d}^{(x)}_{aa}\ket{a^{(x)}}\bra{a^{(x)}}.
\end{align}
This transformation incorporates the polarisation of the optical modes because of the permanent dipoles of the dimer into the definition of the system. 

When carrying out the polaron transformation, the permanent dipoles can introduce a small coherent driving force through the addition of a new self-dipole term which is not taken into account in the secularised Redfield equation. Since only operators that commute with the polaron transformation maintain their original meaning, re-diagonalising the system Hamiltonian leaves us with population states that are ill-defined ~\cite{iles-smith_capturing_2024}. Therefore, we must ensure that the Redfield equation can be applied accurately in the same perturbed basis that the polaron transformation is performed. Since the transition rates are calculated by squaring the interaction Hamiltonian, in order to get any transition-rates terms involving permanent dipoles accurate to the fourth order of dipole coupling it is only necessary to ensure that the Hamiltonian is accurate to the second order.

The self-dipole term that the polaron transformation introduces is equal to \(\lambda( \pmb{d}^{(x)}_{ab}\cdot(\pmb{d}^{(x)}_{aa}+\pmb{d}^{(x)}_{bb}))\). In order to make the off-diagonal elements of the system Hamiltonian cancel with this up to the second order in the dipole coupling, we will use a Schrieffer–Wolff transformation to define the perturbed basis. We begin by applying the unitary transformation S where:
\begin{align}
    S_{ab}=\Bigg{\{} \begin{matrix}
        1 & \text{if } a= b,\\
        \alpha^2 \frac{\lambda( \pmb{d}^{(x)}_{ab}\cdot(\pmb{d}^{(x)}_{aa}+\pmb{d}^{(x)}_{bb}))}{\epsilon_{a}-\epsilon_{b}} & \text{if } a \neq b,
    \end{matrix}
\end{align}
where \(\alpha\) is a dimensionless constant that is used to track the order of the perturbation. This will transform any operators into a system basis that has a one-to-one correspondence with the polaron frame, which we denote with a "bar" and drop the superscript \((x))\). Keeping terms only up to second order in \(\alpha\), 
\begin{align}\label{Eq: System polaron}
    S H^{(x)}_S S^\dagger = \sum_{a} \epsilon_a |\bar{a}\rangle \langle \bar{a}| + \sum_{a \neq b}\alpha \lambda( \pmb{d}^{(x)}_{ab}\cdot(\pmb{d}^{(x)}_{aa}+\pmb{d}^{(x)}_{bb})) |\bar{a}\rangle \langle \bar{b}|\\
    S \tilde{H}_I S^\dagger = \sum_k \sum_{a,b} -if_k \ |\bar{a}\rangle\langle\bar{b}| (\bar{\pmb{d}}_{ab}\cdot \mathbf{e}_k) \hat{a}_k^\dagger +h.c,
\end{align}
where 
\begin{align}
    \bar{\pmb{d}}_{ab} = \Bigg{\{} \begin{matrix}
        \pmb{d}^{(x)}_{ab} & \text{if } a= b,\\
        \pmb{d}^{(x)}_{ab} +\alpha \frac{\lambda( \pmb{d}^{(x)}_{ab}\cdot(\pmb{d}^{(x)}_{aa}+\pmb{d}^{(x)}_{bb}))}{\epsilon_{a}-\epsilon_{b}} (\pmb{d}^{(x)}_{aa}-\pmb{d}^{(x)}_{bb})  & \text{if } a \neq b,
    \end{matrix}
\end{align}
and
\begin{align}
    \ket{\bar{a}} = |a^{(x)}\rangle+ \sum_{b\neq a} \alpha^2 \frac{\lambda( \pmb{d}^{(x)}_{ab}\cdot(\pmb{d}^{(x)}_{aa}+\pmb{d}^{(x)}_{bb}))}{\epsilon_{a}-\epsilon_{b}} |b^{(x)}\rangle.
\end{align}
Note that for any of the original diagonal operators of the system, \(\ket{a^{(x)}}\bra{a^{(x)}}\), after the transformation, all terms that are second order in \(\alpha\) cancel out and hence \(\bar{\mathcal{Z}} = \mathcal{Z}^{(x)}+ O(\alpha^3)\).

We next perform the full polaron transformation on our second-order truncated Hamiltonian. Keeping only terms up to second order in \(\alpha\), the entire Hamiltonian in the polaron frame becomes
\begin{align}
    \bar{H} = \tilde{H}_S- \sum_a\lambda |\pmb{d}^{(x)}_{aa}|^2 |\bar{a}\rangle \langle \bar{a}|+ \sum_k \nu_k a^\dagger_k a_k \\
    +\sum_{a\neq b}  |\bar{a}\rangle \langle \bar{b}|  \bar{C}_{ab}. 
\end{align}
The polaron frame interaction term \(\bar{C}_{ab}\) has an expectation value of zero and is defined as:
\begin{align}
    \bar{C}_{ab} = \sum_k  if_k   (\bar{\pmb{d}}_{ab} \cdot \textbf{e}_k)(a^\dagger_k B(\bar{\pmb{d}}_{aa}-\bar{\pmb{d}}_{bb})- B(\bar{\pmb{d}}_{aa}-\bar{\pmb{d}}_{bb})a_k ),
\end{align}
It should be noted that although \(\bar{C}_{ab}\) itself is not Hermitian, the combination \(\bar{C}_{ab}+ \bar{C}_{ba}\) is, thereby ensuring that the overall Hamiltonian remains Hermitian. Since the Hamiltonian is now diagonalised exactly up to the second order in dipole coupling, we can follow the procedure to find the transition rates outlined in \cite{burgess_strong_2023} while keeping the population states well defined. In doing so, we find the transition rates when accounting for the permanent dipoles in the dimer are
\begin{multline}\label{Eq: Redfield}
     \frac {\partial }{\partial t}\bar{\rho}_{aa}(t)= \sum_{b} \bar{\gamma}_{ab}(\bar{\omega}_{ab})\bar{\rho}_{bb}(t)-\bar{\gamma}_{ab}(-\bar{\omega}_{ab})\bar{\rho}_{aa}(t).
\end{multline}  
The energy differences of the eigenstates in the polaron frame are \(\bar\omega_{ab} = \omega_{ab}- \lambda(|\pmb{d}^{(x)}_{aa}|^2-|\pmb{d}^{(x)}_{bb}|^2) \). The polaron frame transition rates can be expressed by the following Fourier transform.
\begin{multline}
    \bar{\gamma}_{ab}(\bar{\omega}_{ab}) =  \int_{-\infty}^\infty \text {d}s \space \text{e}^{i \bar{\omega} _{ab}s}\text{e}^{ |\bar{\pmb{\Delta}}_{ab}|^2 (\psi_{2}(s)- \psi_{2}(0))} \\
    \times\Big[|\bar{\pmb{d}}_{ab}|^2\psi_{0}(s)
    + |\bar{\pmb{d}}_{ab}\cdot\bar{\pmb{\Delta}}_{ab}|^2\psi_{1}(s)^2\Big],
\end{multline}
where
\begin{align}
    \psi_{n}(s) = \frac{8 \pi}{3}\int_0^\infty \dx \nu \space \frac{J(\nu)}{\nu^n} (\tilde{N}(\nu)\e^{-i \nu s}+(-1)^{n}N(\nu)\e^{i\nu s} ).
\end{align}
By Taylor expanding the exponential of \(\psi_2\) and keeping only the unperturbed terms up to fourth order in the dipole coupling, we find that fourth-order corrections to the transition rates as a result of permanent dipoles in the dimer are
\begin{multline}
    \bar{\gamma}_{ab}(\omega_{ab}) - \gamma_{ab}(\omega_{ab}) =\\
    -( |\pmb{d}^{(x)}_{ab}|^2 |\tilde{\pmb{\Delta}}_{ab}|^2 +|\pmb{d}^{(x)}_{ab}\cdot\tilde{\pmb{\Delta}}_{ab}|^2 ) \mathcal{K}(\omega_{ab}) \\
    - \frac{ |\pmb{d}^{(x)}_{ab}\cdot\pmb{d}^{(x)}_{aa}|^2 - |\pmb{d}^{(x)}_{ab}\cdot\pmb{d}^{(x)}_{bb}|^2 }{\omega_{ab}} 2 \lambda\gamma(\omega_{ab}).
\end{multline}
Here \(\mathcal{K}(\omega_{ab})\) is a strictly positive function that is dependent on the spectral density,
\begin{align}
    \mathcal{K}(\omega_{ab}) = \gamma(\omega_{ab})-\int_{-\infty}^\infty \text {d}s\space \text{e}^{i \omega _{ab}s}\psi_2(s)\psi_0(s).
\end{align}
Since all correction terms are proportional to \(|\pmb{d}^{(x)}_{ab}|^2\), any perfect dark states remain so when accounting for the permanent dipoles of the dimer. The first correction term reduces the transition rate as the change in the dimers' permanent dipoles increases. The contribution of the second term decreases the transition rates between two states when the permanent dipole of the higher-energy state has a larger inner product with the transition dipole than the lower-energy state. Otherwise, this term serves to increase the transition rate, but never more than the amount the first term decreases it. The only mechanism in which permanent dipoles in the dimer can increase the transition rates comes from the changes in \(\omega_{ab}\).

Substituting in a cut-off frequency of \(\lambda = 10\) meV, we find that any correction to the transition rates are approximately of the size  \((10^{-9}eV/\omega_{ab}) \gamma(\omega_{ab})\). This is significantly smaller than other approximations that have been made.

\begin{thebibliography}{10}

\bibitem{rouse_optimal_2019}
D~M Rouse, E~M Gauger, and B~W Lovett.
\newblock Optimal power generation using dark states in dimers strongly coupled to their environment.
\newblock {\em New Journal of Physics}, 21(6):063025, June 2019.

\bibitem{riede_efficient_2011}
Moritz Riede, Christian Uhrich, Johannes Widmer, Ronny Timmreck, David Wynands, Gregor Schwartz, Wolf‐Michael Gnehr, Dirk Hildebrandt, Andre Weiss, Jaehyung Hwang, Sudhakar Sundarraj, Peter Erk, Martin Pfeiffer, and Karl Leo.
\newblock Efficient {Organic} {Tandem} {Solar} {Cells} based on {Small} {Molecules}.
\newblock {\em Advanced Functional Materials}, 21(16):3019--3028, August 2011.

\bibitem{ameri_highly_2013}
Tayebeh Ameri, Ning Li, and Christoph~J. Brabec.
\newblock Highly efficient organic tandem solar cells: a follow up review.
\newblock {\em Energy \& Environmental Science}, 6(8):2390, 2013.

\bibitem{mancal_exciton_2010}
Tomáš Mančal and Leonas Valkunas.
\newblock Exciton dynamics in photosynthetic complexes: excitation by coherent and incoherent light.
\newblock {\em New Journal of Physics}, 12(6):065044, June 2010.

\bibitem{chan_single-photon_2018}
Herman C~H Chan, Omar~E Gamel, Graham~R Fleming, and K~Birgitta Whaley.
\newblock Single-photon absorption by single photosynthetic light-harvesting complexes.
\newblock {\em Journal of Physics B: Atomic, Molecular and Optical Physics}, 51(5):054002, March 2018.

\bibitem{ghosh_decoupling_2024}
Pratyush Ghosh, Antonios~M. Alvertis, Rituparno Chowdhury, Petri Murto, Alexander~J. Gillett, Shengzhi Dong, Alexander~J. Sneyd, Hwan-Hee Cho, Emrys~W. Evans, Bartomeu Monserrat, Feng Li, Christoph Schnedermann, Hugo Bronstein, Richard~H. Friend, and Akshay Rao.
\newblock Decoupling excitons from high-frequency vibrations in organic molecules.
\newblock {\em Nature}, 629(8011):355--362, May 2024.

\bibitem{merrifield_theory_1964}
R.~E. Merrifield.
\newblock Theory of the {Vibrational} {Structure} of {Molecular} {Exciton} {States}.
\newblock {\em The Journal of Chemical Physics}, 40(2):445--450, January 1964.

\bibitem{zhang_dark_2016}
Yiteng Zhang, Aaron Wirthwein, Fahhad~H. Alharbi, Gregory~S. Engel, and Sabre Kais.
\newblock Dark states enhance the photocell power via phononic dissipation.
\newblock {\em Physical Chemistry Chemical Physics}, 18(46):31845--31849, 2016.

\bibitem{fruchtman_photocell_2016}
Amir Fruchtman, Rafael Gómez-Bombarelli, Brendon~W. Lovett, and Erik~M. Gauger.
\newblock Photocell {Optimization} {Using} {Dark} {State} {Protection}.
\newblock {\em Physical Review Letters}, 117(20):203603, November 2016.

\bibitem{zhang_delocalized_2015}
Yiteng Zhang, Sangchul Oh, Fahhad~H. Alharbi, Gregory~S. Engel, and Sabre Kais.
\newblock Delocalized quantum states enhance photocell efficiency.
\newblock {\em Physical Chemistry Chemical Physics}, 17(8):5743--5750, 2015.

\bibitem{mondaini_dark_2018}
S.~V. Kozyrev and I.~V. Volovich.
\newblock Dark {States} in {Quantum} {Photosynthesis}.
\newblock In Rubem~P. Mondaini, editor, {\em Trends in {Biomathematics}: {Modeling}, {Optimization} and {Computational} {Problems}}, pages 13--26. Springer International Publishing, Cham, 2018.

\bibitem{hu_double-excitation_2018}
Zixuan Hu, Gregory~S. Engel, and Sabre Kais.
\newblock Double-excitation manifold's effect on exciton transfer dynamics and the efficiency of coherent light harvesting.
\newblock {\em Physical Chemistry Chemical Physics}, 20(47):30032--30040, 2018.

\bibitem{davidson_eliminating_2022}
Scott Davidson, Felix~A. Pollock, and Erik Gauger.
\newblock Eliminating {Radiative} {Losses} in {Long}-{Range} {Exciton} {Transport}.
\newblock {\em PRX Quantum}, 3(2):020354, June 2022.

\bibitem{davidson_dark_2020}
Scott Davidson, Amir Fruchtman, Felix~A. Pollock, and Erik~M. Gauger.
\newblock The dark side of energy transport along excitonic wires: {On}-site energy barriers facilitate efficient, vibrationally mediated transport through optically dark subspaces.
\newblock {\em The Journal of Chemical Physics}, 153(13):134701, October 2020.

\bibitem{creatore_efficient_2013}
C.~Creatore, M.~A. Parker, S.~Emmott, and A.~W. Chin.
\newblock Efficient {Biologically} {Inspired} {Photocell} {Enhanced} by {Delocalized} {Quantum} {States}.
\newblock {\em Physical Review Letters}, 111(25):253601, December 2013.

\bibitem{garziano_one_2016}
Luigi Garziano, Vincenzo Macrì, Roberto Stassi, Omar Di~Stefano, Franco Nori, and Salvatore Savasta.
\newblock One {Photon} {Can} {Simultaneously} {Excite} {Two} or {More} {Atoms}.
\newblock {\em Physical Review Letters}, 117(4):043601, July 2016.

\bibitem{young_mixed_2020}
Ryan~M. Young and Michael~R. Wasielewski.
\newblock Mixed {Electronic} {States} in {Molecular} {Dimers}: {Connecting} {Singlet} {Fission}, {Excimer} {Formation}, and {Symmetry}-{Breaking} {Charge} {Transfer}.
\newblock {\em Accounts of Chemical Research}, 53(9):1957--1968, September 2020.

\bibitem{hestand_expanded_2018}
Nicholas~J. Hestand and Frank~C. Spano.
\newblock Expanded {Theory} of {H}- and {J}-{Molecular} {Aggregates}: {The} {Effects} of {Vibronic} {Coupling} and {Intermolecular} {Charge} {Transfer}.
\newblock {\em Chemical Reviews}, 118(15):7069--7163, August 2018.

\bibitem{stokes_master_2018}
Adam Stokes and Ahsan Nazir.
\newblock A master equation for strongly interacting dipoles.
\newblock {\em New Journal of Physics}, 20(4):043022, April 2018.

\bibitem{rouse_avoiding_2021}
Dominic~M. Rouse, Brendon~W. Lovett, Erik~M. Gauger, and Niclas Westerberg.
\newblock Avoiding gauge ambiguities in cavity quantum electrodynamics.
\newblock {\em Scientific Reports}, 11(1):4281, February 2021.

\bibitem{breuer_theory_2007}
Heinz-Peter Breuer and Francesco Petruccione.
\newblock {\em The {Theory} of {Open} {Quantum} {Systems}}.
\newblock Oxford University PressOxford, 1 edition, January 2007.

\bibitem{drobizhev_resonance_2007}
M.~Drobizhev, N.~S. Makarov, T.~Hughes, and A.~Rebane.
\newblock Resonance {Enhancement} of {Two}-{Photon} {Absorption} in {Fluorescent} {Proteins}.
\newblock {\em The Journal of Physical Chemistry B}, 111(50):14051--14054, December 2007.

\bibitem{alaraby_salem_two-photon_2015}
M.~Alaraby~Salem and Alex Brown.
\newblock Two-photon absorption of fluorescent protein chromophores incorporating non-canonical amino acids: {TD}-{DFT} screening and classical dynamics.
\newblock {\em Physical Chemistry Chemical Physics}, 17(38):25563--25571, 2015.

\bibitem{deiglmayr_permanent_2010}
J.~Deiglmayr, A.~Grochola, M.~Repp, O.~Dulieu, R.~Wester, and M.~Weidemüller.
\newblock Permanent dipole moment of {LiCs} in the ground state.
\newblock {\em Physical Review A}, 82(3):032503, September 2010.

\bibitem{rogers_investigating_2006}
S.~S. Rogers, P.~Venema, J.~P.~M. Van Der~Ploeg, E.~Van Der~Linden, L.~M.~C. Sagis, and A.~M. Donald.
\newblock Investigating the permanent electric dipole moment of beta‐lactoglobulin fibrils, using transient electric birefringence.
\newblock {\em Biopolymers}, 82(3):241--252, June 2006.

\bibitem{guerout_ground_2010}
R.~Guérout, M.~Aymar, and O.~Dulieu.
\newblock Ground state of the polar alkali-metal-atom–strontium molecules: {Potential} energy curve and permanent dipole moment.
\newblock {\em Physical Review A}, 82(4):042508, October 2010.

\bibitem{jagatap_contributions_2002}
B.~N. Jagatap and William~J. Meath.
\newblock Contributions of permanent dipole moments to molecular multiphoton excitation cross sections.
\newblock {\em Journal of the Optical Society of America B}, 19(11):2673, November 2002.

\bibitem{burgess_strong_2023}
Adam Burgess, Marian Florescu, and Dominic~M. Rouse.
\newblock Strong coupling dynamics of driven quantum systems with permanent dipoles.
\newblock {\em AVS Quantum Science}, 5(3):031402, September 2023.

\bibitem{mirzac_microwave_2021}
Alexandra Mîrzac, Sergiu Carlig, and Mihai~A. Macovei.
\newblock Microwave multiphoton conversion via coherently driven permanent dipole systems.
\newblock {\em Physical Review A}, 103(4):043719, April 2021.

\bibitem{mandal_polarized_2020}
Arkajit Mandal, Sebastian Montillo~Vega, and Pengfei Huo.
\newblock Polarized {Fock} {States} and the {Dynamical} {Casimir} {Effect} in {Molecular} {Cavity} {Quantum} {Electrodynamics}.
\newblock {\em The Journal of Physical Chemistry Letters}, 11(21):9215--9223, November 2020.

\bibitem{shim_permanent_1999}
Moonsub Shim and Philippe Guyot-Sionnest.
\newblock Permanent dipole moment and charges in colloidal semiconductor quantum dots.
\newblock {\em The Journal of Chemical Physics}, 111(15):6955--6964, October 1999.

\bibitem{chestnov_ensemble_2016}
I.~Yu. Chestnov, V.~A. Shakhnazaryan, I.~A. Shelykh, and A.~P. Alodjants.
\newblock Ensemble of asymmetric quantum dots in a cavity as a terahertz laser source.
\newblock {\em JETP Letters}, 104(3):169--174, August 2016.

\bibitem{anton_radiation_2016}
M~A Antón, F~Carreño, O~G Calderón, S~Melle, and E~Cabrera.
\newblock Radiation emission from an asymmetric quantum dot coupled to a plasmonic nanostructure.
\newblock {\em Journal of Optics}, 18(2):025001, February 2016.

\bibitem{chestnov_terahertz_2017}
Igor~Yu. Chestnov, Vanik~A. Shahnazaryan, Alexander~P. Alodjants, and Ivan~A. Shelykh.
\newblock Terahertz {Lasing} in {Ensemble} of {Asymmetric} {Quantum} {Dots}.
\newblock {\em ACS Photonics}, 4(11):2726--2737, November 2017.

\bibitem{warburton_giant_2002}
R.~J. Warburton, C.~Schulhauser, D.~Haft, C.~Schäflein, K.~Karrai, J.~M. Garcia, W.~Schoenfeld, and P.~M. Petroff.
\newblock Giant permanent dipole moments of excitons in semiconductor nanostructures.
\newblock {\em Physical Review B}, 65(11):113303, February 2002.

\bibitem{tretiak_exciton_2000}
Sergei Tretiak, Chris Middleton, Vladimir Chernyak, and Shaul Mukamel.
\newblock Exciton {Hamiltonian} for the {Bacteriochlorophyll} {System} in the {LH2} {Antenna} {Complex} of {Purple} {Bacteria}.
\newblock {\em The Journal of Physical Chemistry B}, 104(18):4519--4528, May 2000.
\newblock Publisher: American Chemical Society (ACS).

\bibitem{fujimoto_electronic_2021}
Kazuhiro~J. Fujimoto.
\newblock Electronic {Couplings} and {Electrostatic} {Interactions} {Behind} the {Light} {Absorption} of {Retinal} {Proteins}.
\newblock {\em Frontiers in Molecular Biosciences}, 8:752700, 2021.

\bibitem{iles-smith_capturing_2024}
Jake Iles-Smith, Owen Diba, and Ahsan Nazir.
\newblock Capturing non-{Markovian} polaron dressing with the master equation formalism.
\newblock {\em The Journal of Chemical Physics}, 161(13):134111, October 2024.

\end{thebibliography}
\end{document}